\documentstyle[prl,aps,multicol,epsf]{revtex}
\input psfig
\begin {document}
\draft
\title {Weiss oscillations in surface acoustic wave propagation}
\author {Y. Levinson}
\address {Condensed Matter Physics, The Weizmann Institute of Science,
Rehovot 76100, Israel}
\author{O. Entin-Wohlman}\address {School of Physics and Astronomy,
Raymond and Beverly Sackler Faculty of Exact Sciences,
\\ Tel Aviv University, Tel Aviv 69978, Israel}
\author { A. D. Mirlin$^*$ and P. W\"olfle}
\address{Institut f\"ur Theorie der kondensierten Materie,
Universit\"at Karlsruhe, 76128 Karlsruhe, Germany}
\date {December 23, 1997}
\maketitle
\begin {abstract}
The interaction of a surface acoustic wave (SAW) with a 
a two-dimensional electron gas in a periodic 
electric potential and a classical magnetic field is considered.
We calculate the attenuation of  the SAW and its velocity change
and show that these quantities exhibit Weiss oscillations.
\end {abstract}
\pacs {PACS numbers: 73.20.Dx, 73.50.Rb}
\begin{multicols}{2}

Oscillations in the  {\it dc} magnetoresistance,
known as Weiss oscillations, can be observed \cite{WKPW89,WKP89,Bet90,PMMS}
 in high mobility two-dimensional electron gas
(2DEG) subject to a perpendicular magnetic field $B$
and a periodic laterally-modulated potential.
These oscillations are geometric in nature, occuring whenever the
cyclotron radius $R_{c}$ is an integral multiple of the
modulation period, $d$.

Surface acoustic waves (SAWs) are known to be a very effective tool
to study the properties of a 2DEG \cite{Wix,Wil,WP}.
Recently the remarkable effect of a periodic potential on the SAW
propagation has been observed in the quantum Hall regime \cite{WWP};
this problem was addressed theoretically in \cite{Adi}. The case of a
strong modulation, when the 2DEG is fragmented into quantum wires was
experimentally studied in \cite{Ben}. In this Letter we focus on a
different aspect: the possibility to observe Weiss oscillations in the
{\it ac} electric field of a SAW in weak (classical) magnetic fields.

We show that when the SAW length  is long compared 
to the grating period  and the electron cyclotron radius
the grating just renormalizes the conductivity $\sigma({\bf q},\omega)$,
where $\omega $ and ${\bf q}$
are the frequency and the wave vector of the SAW.
This  means that the propagation of the SAW in the modulated 2DEG
can be described as propagation in a uniform 2DEG
having an effective conductivity
$\sigma_{eff}({\bf q},\omega)=\sigma({\bf q},\omega)+
\delta\sigma({\bf q},\omega)$, 
where the first term is the conductivity of the 2DEG without grating
and the second is the renormalization due to the grating.
The grating-induced contribution $\delta\sigma$ will be shown 
to exhibit Weiss oscillations, as manifested in the SAW absorption and
the velocity shift.

Due to the piezoelectric properties of GaAs 
the deformation wave is followed by an electrical field wave,
which in the plane of the 2DEG is 
${\bf E}^{0}({\bf r},t)={\bf E}^{0}\exp (-i\omega t+i{\bf q}{\bf r})+c.c.$.
We assume the field to be longitudinal, ${\bf E}^{0}\parallel{\bf q}$,
which corresponds to the usual experimental geometry.
This field is screened by the 2DEG, and because of the modulation
the screened (total) field ${\bf E}({\bf r},t)$
has spatial Fourier components with wave vectors
${\bf q}_{s}={\bf q}+s{\bf p},(s=0,\pm 1,\pm 2,...)$,
where ${\bf p}$ is the wave vector of the grating, $p=2\pi /d$.
It follows that  ${\bf E}({\bf r},t)=
\sum_{s}{\bf E}_{s}\exp (-i\omega t+i{\bf q}_{s}{\bf r})+c.c.$. 
The unscreened field has in these notations only
the $s=0$ Fourier component, ${\bf q}_{0}\equiv {\bf q}$, 
and ${\bf E}_{s}^{0}=\delta _{s,0}{\bf E}^{0}$.
The current density created by the SAW is 
${\bf j}({\bf r},t)=
\sum_{s}{\bf j}_{s}\exp (-i\omega t+i{\bf q}_{s}{\bf r})+c.c.$
and we define
a matrix of tensorial conductivities $\hat{\sigma}_{s,s'}$ by
${\bf j}_{s}=\sum_{s'}\hat{\sigma}_{s,s'}{\bf E}_{s'}$.

Since the SAW velocity $v=\omega /q$ 
(equal to $2.8\times 10^{5}$cm/sec in GaAs)
is much less than the light velocity one can neglect retardation
effects and find the screening field solving a quasistatic problem.
The result is (a detailed account of our work will be given
elsewhere \cite{long}) 
\begin{equation}
{\bf E}_{s}-{\bf E}_{s}^{0}=-{\frac{2\pi i}{\omega {\epsilon }_{o}}}
{\frac{{\bf q}_{s}}{q_{s}}}\left( {\bf q}_{s}\cdot \sum_{s'}
\bbox{\hat{\sigma}}_{s,s'}\cdot {\bf E}_{s'}\right),
\label{scr}
\end{equation}
where ${\epsilon }_{o}$ is the effective dielectric constant of the
background.

It follows from this equation that when ${\bf E}^{0}\parallel {\bf q}$,
as assumed, then for 
all $s$ one has ${\bf E}_{s}\parallel {\bf q}_{s}$, i.e.
the total screened field is also longitudinal.
In this case one can put 
${\bf E}_{s}=({\bf q}_{s}/q_{s})E_{s}$
and eliminate the tensorial properties, obtaining from Eq. (\ref{scr}) 
the relation between the screened and unscreened fields 
in a scalar form
$\sum_{s'}{\epsilon }_{s,s'}E_{s'}=\delta_{s,0}E^{0}$,
where the longitudinal dielectric constant matrix is 
${\epsilon }_{s,s'}=\delta _{s,s'}+(2\pi i/
\omega \epsilon_{o})q_{s}{\sigma }_{s,s'} $
and the longitudinal conductivity matrix is 
$\sigma _{s,s'}=({\bf q}_{s}/q_{s})\cdot
 \bbox{\hat{\sigma}}_{s,s'}\cdot ({\bf q}_{s'}/q_{s'})$.

The absorption $Q$ of the SAW  is given by a spatial and temporal
average of ${\bf j}\cdot {\bf E}$. Using the field and current
representations one finds for longitudinal fields
$Q=\sum_{s,s'}E_{s}^{*}{\sigma }_{s,s'}E_{s'}+c.c..$
Expressing the screened field ${\bf E}$ in terms of the SAW field ${\bf
E}^{0} $ one has  
\begin{equation}
\label{Q}
Q/|{\bf E}^{0}|^{2}=i{\sigma }_{M}(\epsilon ^{-1})_{0,0}+c.c.
=-2\sigma_{M}\Im(\epsilon ^{-1})_{0,0}. 
\end{equation}
Here $(\epsilon^{-1} )_{0,0}$ is the $s=0,s'=0$ matrix
element of the matrix inverse to $\epsilon$ and $\sigma _{M}=
\epsilon _{o}v/ 2\pi $ (equal to $ 1\times 10^{-2}(e^{2}/h )$
for GaAs). 

Furthermore, it can be shown \cite{long} that in the lowest order of 
the piezoelectric coupling constant $\alpha$ (in GaAs $\alpha^2/2=3.2\times
10^{-4}$) the renormalized dispersion relation of the SAW takes the form
\begin{equation}
\label{dr}
\omega=v q\left[1+(\alpha^2/2)(\epsilon^{-1})_{0,0}\right].
\end{equation}
As a consequence, both  the relative shift of the SAW velocity 
$\Delta v/v$ and  
its attenuation coefficient per unit length $\Gamma$ are given 
in terms of $(\epsilon ^{-1})_{0,0}$ as follows
\begin{eqnarray}
&& \Delta v/v=(\alpha^2/ 2)\Re (\epsilon^{-1})_{0,0}\ ;
\label{dv}\\
&& \Gamma =-q(\alpha^2/ 2)\Im (\epsilon^{-1})_{0,0}.
\label{gamma}
\end{eqnarray}

For the calculation of the longitudinal conductivities
which define $(\epsilon^{-1})_{0,0}$ we
 start with the kinetic equation 
for the distribution function $f({\bf r}, {\bf k},t)$
of the 2DEG electrons in a form used in \cite{B}
\begin{eqnarray}
&&\left[ {\frac{\partial}{\partial t}}+{\bf v}\nabla+ {\frac{e}{c}}{\bf v}
\times {\bf B}{\frac{\partial}{\partial {\bf k} }}+ (e{\bf E}
({\bf r},t) -\nabla U({\bf r}))
{\frac{\partial}{\partial {\bf k} }}\right]f \nonumber \\
&&\hspace{3cm} =-(f-\langle f\rangle)/\tau. 
\label{keq}
\end{eqnarray}
Here, ${\bf v}={\bf k}/m$, $U({\bf r})$ denotes the modulating
potential in the plane of the 2DEG, ${\bf E}({\bf r},t)$ is the
total electric field of the SAW, $\tau $ is the momentum
relaxation time (assuming short range scattering potential),
the angular brackets  $\langle\ldots\rangle$ denote 
an average over the directions of ${\bf k}$, and $e=-|e|$ is the
electron charge.

When ${\bf E}({\bf r},t)=0$ the 2DEG is in equilibrium and the solution of
Eq. (\ref{keq}) is
$ f_{0}({\bf r},{\bf k})=f_{T}({\varepsilon }_{k}+U({\bf r}))$,
where ${\varepsilon }_{k}=k^{2}/2m$ and $f_{T}({\varepsilon })$ is the Fermi
distribution.
To linearize Eq. (\ref{keq}) with respect to 
${\bf E}={\bf E}({\bf r})\exp (-i\omega t)+c.c.$ we write
$f({\bf r},{\bf k},t)=f_{0}({\bf r},{\bf k})+\delta f({\bf r},{\bf k},t)$
with 
$\delta f({\bf r},{\bf k},t)=\exp (-i\omega t)(-{\frac{\partial }{\partial 
\varepsilon }}f_{0}({\bf r},{\bf k}))F({\bf r},{\bf k})+c.c.$.

For our purposes, it is sufficient to consider
the zero temperature case, for which
$(-{\frac{\partial }{\partial \varepsilon }}f_{0}({\bf r},{\bf k}))=
\delta (\varepsilon _{k}+U({\bf r})-\varepsilon _{F})$,  
where $\varepsilon _{F}={1\over 2}mv_{F}^2$
is the Fermi energy. Consequently the magnitude of
the electron velocity $v({\bf r})$ is defined by the grating potential
according to 
${\frac{1}{2}}mv({\bf r})^{2}+U({\bf r})={\varepsilon }_{F}$
and  $F({\bf r},{\bf k})$ becomes a function $F({\bf r},{\bf n})$ of the
unit vector ${\bf n}={\bf v}/v({\bf r})$.
The current density (taking into account the two spin orientations)
is ${\bf j}({\bf r})=e(m/\pi )v({\bf r})
\langle{\bf n}F({\bf r},{\bf n})\rangle$.

After linearization with respect to the electric field ${\bf E}$ one obtains
a linear equation for $F$, 
\begin{equation}
\label{leq}
LF({\bf r},{\bf n})=v({\bf r})e{\bf E}({\bf r}){\bf n,} 
 \label{LF}
\end{equation}
where the operator $L$ is
$$
L \equiv -i\omega +\frac{1}{\tau}+v({\bf r}){\bf n}\nabla
 +({\bf e}\nabla v({\bf r})
+\omega _{c})\frac{\partial }{\partial \varphi }
-\frac{1}{\tau }\int \frac{d\varphi }{2\pi }.
$$
Here the angle $\varphi $ defines the direction of ${\bf n}$, 
$\omega_{c}=|e|B/mc$ is the cyclotron frequency and ${\bf e}
={\bf n}\times {\bf B}/B$.
To calculate the tensorial conductivity ${\hat{\sigma}}_{s',s}$ we
put ${\bf E}({\bf r})={\bf E}_{s}\exp(i{\bf q}_{s}{\bf r})$ 
and represent
$F({\bf r},{\bf n})=\sum_{s'}\chi _{s',s}({\bf n})
\exp (i{\bf q}_{s'}{\bf r})$.

Consider now a periodic potential $U({\bf r})=U_{0}\cos({\bf p}{\bf r})$.
In the case of weak modulation $\eta\equiv U_{0}/{\varepsilon}_{F}\ll 1$
and one can perform a systematic expansion of Eq. (\ref{keq}) in $\eta$.
This implies expanding the functions 
$v({\bf r})$ and $\chi _{s',s}({\bf n})$
and comparing in the resulting equation
terms with the same spatial dependence and of the
same order in $\eta $.
One then finds that since the grating has only
the first harmonic in ${\bf p}$, the non-vanishing
components of the tensorial conductivity are:
$(s,s)$ to order $\eta^{0}$,
$(s\pm 1,s)$ to order $\eta^{1}$, and $(s,s)$, $(s\pm 2,s)$
to order $\eta^{2}$.

The explicit expressions for the longitudinal conductivities are
\begin{eqnarray}
\sigma _{s,s}^{(0)} &=& 2\sigma _{0}\frac{\omega ^{2}}
{v_{F}^{2}q_{s}^{2}}\left[ {\frac{1}{i\omega \tau }}+
\langle G_{s}\rangle\right] , 
\label{s0} \\
\sigma _{s\pm 1,s}^{(1)} &=& \eta \sigma _{0}\frac{\omega ^{2}}
{2v_{F}^{2}q_{s}q_{s\pm 1}}iv_{F}\tau 
\langle G_{s\pm 1}d_{s}^{\pm 1}G_{s}\rangle,  
\label{s1} \\
\sigma_{s,s}^{(2)} &=& \eta ^{2}\sigma _{0}\frac{\omega ^{2}}
{8v_{F}^{2}q_{s}^{2}} (iv_{F}\tau)^{2}
\nonumber\\
\label{s2}
&\times& \langle G_{s}d_{s+1}^{-1}G_{s+1}d_{s}^{+1}G_{s}+
G_{s}d_{s-1}^{+1}G_{s-1}d_{s}^{-1}G_{s} \rangle.
\end{eqnarray}
(As is shown below,
$(\epsilon^{-1})_{0,0}$ does not require the $(s\pm 2,s)$  elements
of $\sigma^{(2)}$. To simplify the results the Fermi velocity is renormalized
according to $v_{F}\rightarrow v_{F}(1-\eta^{2}/16)$).

Here 
$\sigma _{0}=(m/2\pi )e^{2}v_{F}^{2}\tau$
is the {\it dc} conductivity of a homogeneous 2DEG when $B=0$ 
and  the following operators are introduced:
$d_{s}^{k}
={\bf n}{\bf q}_{s}+k{\bf p}{\bf e}\partial /\partial \varphi$ ,
\begin{equation}
\label{G}
G_{s}=\left[ 1-i\omega \tau +iv_{F}\tau {\bf n}
{\bf q}_{s}+\omega _{c}\tau \frac{\partial }{\partial \varphi }
-\int \frac{d \varphi }{2\pi }\right] ^{-1}. 
\end{equation}

The explicit calculation of the conductivity
can be performed in terms
of the integral representation of the operator $G_{s}$
given by 
\begin{equation}
\label{R}
G_{s}g(\varphi)=R_{s}g(\varphi )+{\frac{\langle R_{s}g\rangle}
{1-\langle R_{s}\rangle}}R_{s},
\end{equation}
where the operator $R_{s}$ is defined by
\begin{equation}
\label{K}
R_{s}g(\varphi )=\gamma \int_{-\infty }^{\varphi } d\varphi'
e^{K_{s}(\varphi,\varphi')}g(\varphi'), 
\end{equation}
with 
$
K_{s}(\varphi,\varphi')=-\nu (\varphi -\varphi')
-iz_{s}(\sin \varphi -\sin \varphi')
$
and $\nu =(\tau ^{-1}-i\omega )/\omega _{c}$,
 $\gamma =1/\omega _{c}\tau $ and $z_{s}=q_{s}v_{F}/\omega _{c}$.
The angles $\varphi $ and $\varphi'$ are counted from the
direction of ${\bf q}_{s}$ and $R_{s}$, $\langle R_{s}\rangle$
stand for $R_{s}1$ and $\langle R_{s}1\rangle$, respectively.

In what follows we consider a "fast" grating with a period shorter than the 
SAW length $\lambda$, i.e. $ q\ll p$.
(As an example, for $\omega /2\pi =300$MHz one has $\lambda=9\mu$m
while $d$ varies from $0.1\mu$m to $1\mu$m).
In this case one can present the result of inverting
the matrix $\epsilon$ in the following way
\begin{eqnarray}
(\epsilon^{-1})_{0,0}=
\frac{1}{1+i\sigma_{eff}({\bf q},\omega)/\sigma_{M}}
\label{e}
\end{eqnarray}
with $\sigma_{eff}({\bf q},\omega)=\sigma(q,\omega)+
\delta\sigma({\bf q},\omega)$,
where $\sigma(q,\omega) $ is the longitudinal
conductivity of a homogeneous 2DEG  corresponding to
wave vector $q$ and frequency $\omega $ and
\begin{eqnarray}
&&\delta\sigma({\bf q},\omega) \nonumber \\
&& =\sigma_{0,0}^{(2)} 
-\frac{\sigma_{0,1}^{(1)}\sigma_{1,0}^{(1)}}
{\sigma_{1,1}^{(0)}-i(q_{0}/q_{1})\sigma_{M}}
-\frac{\sigma_{0,\bar{1}}^{(1)}\sigma_{\bar{1},0}^{(1)}}
{\sigma_{\bar{1},\bar{1}}^{(0)}-i(q_{0}/q_{\bar{1}})\sigma_{M}}. 
\label{dels}
\end{eqnarray}
(Here $\bar{1}\equiv -1$).
The result (\ref{dels}) for $\delta\sigma$ is valid up to the second
order in $\eta$, assuming $\eta$ to be small. We will show below that
nonetheless in the case ${\bf q}\bot{\bf p}$ and for 
$\omega_c\tau\gg 1$  $\delta\sigma$ can be comparable to $\sigma$,
since
$\delta\sigma/\sigma\propto\eta^2(\omega_c\tau)^2$. Eq. (\ref{dels}) is
however 
correct even in this case, since the higher order terms are small, of
order  $\eta^2$, compared to $\delta\sigma$. 

Note that for a fast grating  we have approximately $q_{1}=q_{{\bar{1}}}=p$
and  $\sigma_{1,1}^{(0)}=\sigma_{{\bar{1}},{\bar{1}}}^{(0)}=
\sigma(p,\omega)$, while  $\sigma_{0,0}^{(0)}=\sigma(q,\omega)$.
As a result one can simplify Eq. (\ref{dels}) to obtain
\begin{eqnarray}
\delta\sigma({\bf q},\omega)=
\sigma_{0,0}^{(2)}
-\frac{\sigma_{0,1}^{(1)}\sigma_{1,0}^{(1)}+
\sigma_{0,\bar{1}}^{(1)}\sigma_{\bar{1},0}^{(1)}}
{\sigma(p,\omega)-i(q/p)\sigma_{M}}.
\label{dels1}
\end{eqnarray}

The conductivities of a homogeneous 2DEG are given by \cite{MW}
\begin{equation}
\label{sq}
\sigma(q_{s},\omega)\equiv\sigma_{s,s}^{(0)}=
2\sigma _{0}\displaystyle\frac{\omega ^{2}}{v_{F}^{2}q_{s}^{2}}
\left[ {\frac{1}{\ i\omega \tau }}+\displaystyle\frac{\langle R_{s}\rangle}
{1-\langle R_{s}\rangle}\right],  
\end{equation}
where
\begin{equation}
\label{JJ}
\langle R_{s}\rangle=(\pi\gamma /\sinh \pi \nu )
J_{i\nu}(z_{s})J_{-i\nu}(z_{s})
\end{equation}
and $J_{\pm i\nu}$ are the Bessel functions.

We assume in addition
that $\lambda$ is longer than the 
cyclotron radius $R_{c}=v_{F}/\omega _{c}$, i.e. $z_{0}=qv_{F}/\omega
_{c}=qR_{c}\ll 1$.
(For $B=0.1$T and $v_{F}=1.3\times 10^{7}$cm/sec the
cyclotron radius $R_{c}=0.5\mu$m, and for  
$\omega /2\pi=300$MHz one finds $qR_{c}=0.3$).
Since the Weiss oscillations take place at $pR_{c}\sim
\pi n$ with $n=1,2,\ldots$, the above assumption $p\gg q$ implies that
a considerable number of oscillations belongs to the region $qR_c\ll 1$.

In the case $qR_{c}\ll 1$ one finds from Eq. (\ref{sq}) 
 the longitudinal conductivity of the homogeneous 2DEG to be 
\begin{equation}
\label{sqs}
\sigma(q,\omega )=\sigma_{0}
\frac{1-i\omega\tau}
{(1-i\omega\tau)^2+(\omega_{c}\tau)^2+i(qv_{F}\tau)^2/2\omega\tau},
\end{equation}
while the correction $\delta\sigma$, Eq. (\ref{dels1}), is reduced to the
 form 
\begin{eqnarray}
\delta\sigma({\bf q},\omega)= \eta^2 \sigma_{0}
 \displaystyle\frac{(\omega\tau)^2}{8(1-\langle R_{0}\rangle)^2}
 [\Phi_{2}-\xi\Phi_{1}].
\label{dels2}
\end{eqnarray}
The functions $\Phi_{1,2}$  depend on $\gamma$, $\nu$,
$z=pR_{c}=2\pi R_{c}/d $ and on the propagation direction of the SAW
given by the angle $\theta$ between ${\bf q} $ and ${\bf p}$,
\begin{eqnarray}
\Phi_{1}&=&2\frac{\gamma^2}{(\nu ^2+1)^2}
\left[\frac{z\langle R_{1}\rangle}{1-\langle R_{1}\rangle}\right]^2 
D_{\nu}(\theta), \label{P1}\\
\Phi_{2}&=&-\frac{\gamma\nu }{\nu ^2+1}+
\frac{\nu^2 }{(\nu ^2+1)^2}
\frac{2z^2\langle R_{1}\rangle }{1-\langle R_{1}\rangle}D_{\nu}(\theta),
\label{P2}
\end{eqnarray}
where
\begin{eqnarray}
D_{\nu}(\theta)=\cos^2\theta-\nu ^{-2}\sin^2\theta.
\label{D}
\end{eqnarray}
~From Eqs. (\ref{P1}), (\ref{P2}) and (\ref{D})
it follows that when ${\bf q}\perp {\bf p}$
the grating induced part of the conductivity 
contains in addition to the small factor $\eta^2$
also a  factor $(\omega_{c}\tau)^2$ which can be large.

The function  $\xi$ is independent of the propagation direction
\begin{equation}
\label{xi}
\xi^{-1}=(1-i\omega\tau)
\Bigl [\frac{\langle R_{1}\rangle }{1-\langle R_{1}\rangle}
+\frac{1+pa_{B}/2}{i\omega\tau}\Bigr],
\end{equation}
where $a_{B}=\epsilon _{o}/me^{2}$ is the Bohr radius (equal $100\AA $ for
GaAs).
Using Eq. (\ref{JJ}) to calculate $\langle R_{0}\rangle$ and
$\langle R_{1}\rangle$ in 
the functions $\Phi_{1,2}$ and $\xi$
one can find $\delta\sigma({\bf q},\omega)$ and the SAW propagation
properties from Eqs. (\ref{dv}) and (\ref{gamma}).

The Weiss oscillations are expected to be visible
when $\omega _{c}\tau \gg 1 $ and $z=pR_{c}\agt 1$.
(For  $B=0.1$T and $\tau=$150 ps
one finds  $\omega _{c}\tau =40$).
We assume in addition that $\omega \tau \ll 1$.
(For  $\omega /2\pi=300$MHz 
with  $\tau=$150 ps  one finds $\omega \tau =0.3$).
With these approximations (when $\gamma=\nu=(\omega_{c}\tau)^{-1}\ll 1$)
the results are more transparent.

The longitudinal conductivity with no grating is given now by
\begin{eqnarray}
\label{sqss}
\sigma(q,\omega )=
\frac{\sigma_{0}}{(\omega_{c}\tau)^2}
\left[1+{i(qR_{c})^{2}\over 2\omega\tau}\right]^{-1}.
\end{eqnarray}
Note that $\sigma_{0}/(\omega_{c}\tau)^{2}$ is the
"{\it dc} conductivity" at strong magnetic fields,
which is obtained upon solving the kinetic equation
with an electric field constant in time and homogeneous
in space. This "{\it dc} conductivity"
corresponds to the condition $(qR_{c})^{2}/2\omega \tau \ll 1$,
which is not always satisfied in SAW measurements.

In the assumed range of parameters, the term $\xi\Phi_{1}$ in 
Eq. (\ref{dels2}) can be neglected compared to
$\Phi_{2}$, yielding
\begin{eqnarray}
\delta\sigma({\bf q},\omega)&=&\sigma_{0,0}^{(2)}=
-\frac{\eta^2\sigma_0}{8(\omega _{c}\tau )^{2}}
\left[ 1+i{\frac{(qR_{c})^{2}}{2\omega \tau }}\right] ^{-2}
\nonumber\\
&\times&
\{-1+\phi(z)
 \left[\cos ^2\theta - (\omega_{c}\tau)^2 \sin ^2\theta\right]\}.
\label{ds}
\end{eqnarray}
with
\begin{eqnarray}
\phi(z)=2z^{2}J_{0}^{2}(z)/[1-J_{0}^{2}(z)].
\end{eqnarray}
These results demonstrate that the effect of the grating
and hence also the amplitude of the Weiss oscillations is stronger
by a factor of $(\omega_{c}\tau)^2$ when 
the SAW propagates perpendicular to the grating,
${\bf E}^{0}\parallel{\bf q}\perp {\bf p}$, compared to the case of parallel
propagation, ${\bf E}^{0}\parallel{\bf q}\parallel {\bf p}$.
This is because in the ``perpendicular'' geometry the grating
affects the Hall current, which is stronger.
In the limit $q\to 0$ Eq. (\ref{ds}) reduces to the result for the {\it
dc} magnetoconductivity derived in \cite{B}. 

\begin{figure}
\narrowtext
\centerline{\psfig{figure=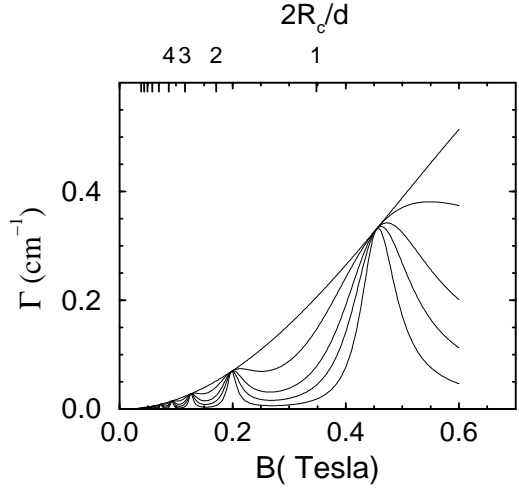,width=9cm}}
\vspace{-0.5cm}
\caption{Absorption coefficient $\Gamma$ of SAW as a function of the
magnetic field for the transverse orientation, ${\bf p}\bot{\bf q}$ and
the following values of the parameters: density $n_e=1\cdot
10^{11} {\rm cm}^{-2}$, $\omega=2\pi\cdot 300 {\rm MHz}$, $\tau=150 {\rm ps}$,
$d=300 {\rm nm}$.
The strength of the grating $\eta$ is equal to (from the top to the
bottom): 0 (no grating), 0.01, 0.02, 0.03, and 0.05.}
\label{fig1}
\end{figure}

\vspace{-0.5cm}

\begin{figure}
\narrowtext
\centerline{\psfig{figure=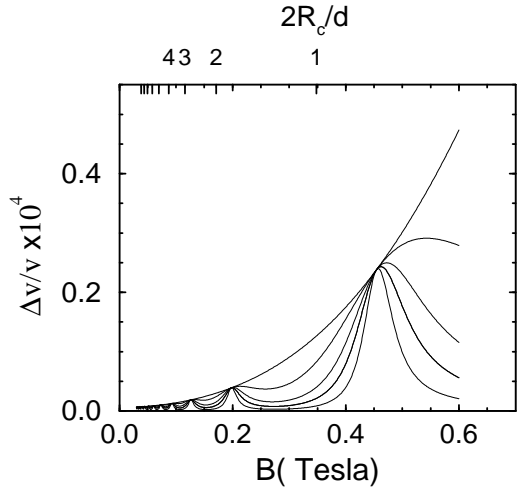,width=9cm}}
\vspace{-0.5cm}
\caption{Velocity shift $\Delta v/v$ of SAW as a function of the
magnetic field. The values of all the parameters are the same as in Fig.1.}
\label{fig2}
\end{figure}

We have evaluated numerically the attenuation and the velocity change
according to Eqs. (\ref{dv}), (\ref{gamma}), (\ref{e}), and
(\ref{dels2}) using typical experimental parameters in the regime
$d,R_c\ll\lambda$. The results are shown in Figs. 1 and 2 for the
orientation ${\bf q}\bot{\bf p}$. In this case the oscillations
amplitude is proportional to $(\omega_c\tau)^2\eta^2$ (see
Eq. (\ref{ds}) for $\theta=\pi/2$), which can be large even for small
$\eta$. This is demonstrated in the figures, where $\eta=0.01$ is
sufficient to produce strong oscillations ($\omega_c\tau$ is as large
as 100 at $B=0.25 {\rm T}$). For the parallel orientation ${\bf
q}\parallel{\bf p}$ ($\theta=0$) the factor of $(\omega_c\tau)^2$ is
absent in (\ref{ds}) and the amplitude of oscillations is proportional
to $\eta^2$ and thus small for weak modulation.

Let us mention that in high mobility samples the impurity potential is
smoothly varying and the scattering probability is not isotropic, but
rather peaked in the forward direction. This will strongly affect the
damping of the oscillations with high oscillation number 
$2R_c/d\gg 1$, but will not be very important for the first few
oscillations with $2R_c/d\sim 1$.

We thank A.~Stern and F.~von~Oppen for discussions of unpublished
results. 
This work was supported by the Alexander von Humboldt Foundation
(Y.~L.), the Albert Einstein Minerva Center for Theoretical Physics
(O.~E.-W.), the SFB195 der Deutschen
Forschungsgemeinschaft (A.~D.~M. and P.~W.), and the German--Israeli
Foundation.

\end{multicols}
\end {document}